\documentclass[conference]{IEEEtran}
\IEEEoverridecommandlockouts
\usepackage{cite}
\usepackage{overpic}
\usepackage{amsmath,amssymb,amsfonts}
\usepackage{graphicx}
\usepackage{textcomp}
\usepackage{xcolor}
\usepackage{float}
\usepackage{amsthm}
\usepackage{graphicx}
\usepackage{epstopdf}
\usepackage{amsmath,bm}
\usepackage{amsfonts}
\usepackage{amssymb}
\usepackage{color}
\usepackage{subfigure}
\usepackage{multirow}
\usepackage{multicol}
\usepackage{url}
\usepackage{soul,xcolor}
\usepackage{algorithm}
\usepackage{algpseudocode}%

\theoremstyle{plain}

\newtheorem{lemma}{Lemma}

\newcommand{\vect}[1]{\mathbf{#1}}

\def\Htran{\mbox{\tiny $\mathrm{H}$}}
\def\Ttran{\mbox{\tiny $\mathrm{T}$}}
\def\CN{\mathcal{N}_{\mathbb{C}}} 

\begin{document}

\title{Fair and Energy-Efficient Activation Control Mechanisms for Repeater-Assisted Massive MIMO
\thanks{This work has been funded by Celtic-Next project RAI-6Green partly supported by Swedish funding agency Vinnova and SSF SUCCESS project.
}
}

\author{\IEEEauthorblockN{Ozan Alp Topal\IEEEauthorrefmark{1}, Özlem Tuğfe Demir\IEEEauthorrefmark{2}, Emil Björnson\IEEEauthorrefmark{1}, and Cicek Cavdar\IEEEauthorrefmark{1}}
\IEEEauthorblockA{ {\IEEEauthorrefmark{1}Department of Computer Science, KTH Royal Institute of Technology, Kista, Sweden
		} \\
  {\IEEEauthorrefmark{2}Department of Electrical-Electronics Engineering, TOBB University of Economics and Technology, Ankara, Turkey
		} \\
		\IEEEauthorblockA{E-mail: \IEEEauthorrefmark{1}\{oatopal, emilbjo, cavdar\}@kth.se, \IEEEauthorrefmark{2}ozlemtugfedemir@etu.edu.tr}
}
}

\maketitle

\begin{abstract}
Massive multiple-input multiple-output (mMIMO) has been the core of 5G due to its ability to improve spectral efficiency and spatial multiplexing significantly; however, cell-edge users still experience performance degradation due to inter-cell interference and uneven signal distribution. While cell-free mMIMO (cfmMIMO) addresses this issue by providing uniform coverage through distributed antennas, it requires significantly more deployment cost due to the fronthaul and tight synchronization requirements. Alternatively, repeater-assisted massive MIMO (RA-MIMO) has recently been proposed to extend the coverage of cellular mMIMO by densely deploying low-cost single-antenna repeaters capable of amplifying and forwarding signals. In this work, we investigate amplification control for the repeaters for two different goals: (i) providing a fair performance among users, and (ii) reducing the extra energy consumption by the deployed repeaters. We propose a max-min amplification control algorithm using the convex-concave procedure for fairness and a joint sleep mode and amplification control algorithm for energy efficiency, comparing long- and short-term strategies. Numerical results show that RA-MIMO, with maximum amplification, improves signal-to-interference-plus-noise ratio (SINR) by over  $20\,\text{dB}$  compared to mMIMO and performs within $1\,\text{dB}$ of cfmMIMO when deploying the same number of repeaters as access points in cfmMIMO. Additionally, our majority-rule-based long-term sleep mechanism reduces repeater power consumption by  $70\%$ while maintaining less than $1\%$  spectral efficiency outage.  
\end{abstract}


\section{Introduction}

Massive multiple-input multiple-output (mMIMO) has been a cornerstone of 5G networks, leveraging large antenna arrays to enhance spectral efficiency (SE) and spatial multiplexing. By exploiting advanced beamforming techniques, mMIMO can serve multiple users simultaneously within the same time-frequency resources, significantly improving network capacity \cite{massivemimobook}. However, despite these advantages, cell-edge users continue to experience performance degradation due to inter-cell interference and uneven signal distribution, limiting the overall network fairness. To address this issue, distributed MIMO approaches have been intensively analyzed, including cell-free mMIMO (cfmMIMO) to eliminate the concept of traditional cellular boundaries by deploying a large number of distributed antennas that jointly serve all users. This architecture offers uniform coverage and reduces inter-cell interference, making it a promising alternative to conventional mMIMO. However, implementing cfmMIMO at scale requires extensive fronthaul infrastructure and tight synchronization among distributed access points, leading to high deployment and operational costs \cite{topal2024energy}.

 Reconfigurable intelligent surface (RIS) has also gained significant attention as a means to enhance wireless communication by passively reflecting signals to improve coverage and energy efficiency \cite{RIS_magazine}. Unlike active beamforming in mMIMO, RIS uses tunable meta-atoms to intelligently alter the propagation environment without requiring additional power-consuming radio-frequency chains. Nevertheless, RIS deployment presents notable challenges, including the need for continuous control signaling and complicated channel state information (CSI) acquisition algorithms, which significantly increase computational overhead and implementation costs \cite{aastrom2024ris}. These limitations motivate the exploration of alternative, cost-effective solutions to extend the coverage of mMIMO systems.

Alternatively, low-cost network-controlled repeaters have been widely used for network coverage extension, where they are also a part of Release 18 \cite{wen2024shaping}. Coverage enhancements have been studied considering line-of-sight (LOS) environments \cite{LOS_repeater}, and  in some practical applications such as railway networks \cite{repeater_power_consumption}. In \cite{erik_larsson}, the authors envision using the low-cost repeaters not to extend the coverage of a network, but to be deployed within the
cell to increase macro diversity as in cfmMIMO by acting as active scatterers with amplification. The authors named such a system as \emph{repeater-assisted massive MIMO (RA-MIMO)}, and provided the requirements of such repeaters, showing that the performance of these systems could approach that of cfmMIMO. As a difference from widely used relays, the repeaters in this context are capable of instantaneously amplifying and forwarding the re-transmitting signal within hundreds of nanoseconds \cite{erik_larsson}.
Later, the stability conditions for repeaters were analyzed in \cite{larsson2024stability}, demonstrating that the amplification at repeaters is upper-bounded by deployment constraints and inter-repeater distance to prevent infinite amplification loops, which could otherwise lead to delay and interference. Although RA-MIMO provides a practical means of enhancing user performance, optimizing repeater amplification control is crucial to balance fairness among users and minimize excessive energy consumption.

In this work, we formulate and analyze amplification control strategies for RA-MIMO under two key objectives: (i) ensuring fair performance among users and (ii) minimizing additional power consumption. We propose a max-min amplification control algorithm using the convex-concave procedure (CCP) to address fairness and a joint sleep mode and amplification control algorithm to enhance energy efficiency. Our results demonstrate that RA-MIMO significantly improves signal-to-interference-plus-noise ratio (SINR) compared to conventional mMIMO and closely matches cfmMIMO’s performance when deploying an equal number of repeaters as access points. Furthermore, our proposed long-term sleep mechanism reduces power consumption by $70\%$ compared to the scenario where all repeaters remain active and transmit at maximum power, while maintaining minimal SE outage. This highlights RA-MIMO’s potential as an efficient and scalable solution for next-generation wireless networks.

\section{System Model}
We consider the uplink of a repeater-assisted mMIMO (RA-MIMO) system consisting of an $M$-antenna mMIMO base station (BS), $K$ single-antenna user equipments (UEs), and $L$ single-antenna repeaters. The repeaters can be arbitrarily located in the considered coverage area. As in \cite{erik_larsson}, for simplicity, we will disregard the interactions between the repeaters. The received signal at repeater $l$ is given by
\begin{equation}
    \tilde{r}_l = \sqrt{\rho_u} \sum_{i=1}^K h_{l,i} s_i + n_l, 
\end{equation}
where $h_{l,i} \in \mathbb{C}$ is the channel coefficient between UE $i$ and repeater $l$, $s_i \in \mathbb{C}$ is the uplink transmit signal of UE $i$, where $\mathbb{E}[|s_i|^2]=1$. Moreover, ${\rho_u}$ is the uplink power by the UEs, which is assumed to be equal and set to the maximum in this work. The additive white Gaussian noise (AWGN) is denoted by $n_l \sim \CN(0, \sigma_{\rm r}^2)$. The repeater $l$ will amplify and transmit the received signal. Hence, the signal transmitted by repeater $l$ becomes
\begin{equation}
    r_l = \alpha_l  \tilde{r}_l \triangleq \sqrt{\rho_u} \alpha_l\sum_{i=1}^K h_{l,i} s_i + \alpha_l n_l, 
\end{equation}
where $\alpha_l \in \mathbb{R}$ is the amplification factor. It satisfies $0\leq\alpha_l \leq \alpha_{\mathrm{max}} $, where $\alpha_{\mathrm{max}}$ is the maximum amplification factor that depends on both the hardware limitations and the inter-repeater distance to guarantee stability\footnote{Stability in this context guarantees that the repeaters do not enter an infinite amplification loop of the signals received from each other.}. We assume the LOS link dominates inter-repeater channels and follow the stability analysis in \cite{larsson2024stability}. This limit also constitutes a lower bound on other channels as well. Another upper limit for the repeater is the output power, which can be calculated as 
\begin{equation}
   P_{\mathrm{out},l} = \alpha^2_l \left(\rho_u \|\vect{h}_{l}\|^2  + \sigma^2_{\rm r} \right)  \leq P_{\mathrm{max}},
\end{equation}
where $\vect{h}_{l} = [h_{l,1}, \ldots, h_{l,K}]^{\Ttran}  $.
Note that the repeater amplifies the noise in addition to the received signal, which will be discussed as a potential problem in the following sections.  The received signal by the BS is the combination of the signals received through the direct channels from the UEs and the amplified signals from the repeaters. As detailed in \cite{erik_larsson}, the delay of this operation is negligible, and the received signal can be represented as 
\begin{align}
    \vect{y} &=  \sum_{l=1}^L \vect{g}_l r_l + \sqrt{\rho_u}\sum_{i=1}^K \overline{\vect{h}}_i s_i + \vect{v},   \\
    & \triangleq \sqrt{\rho_u}  \sum_{l=1}^L \sum_{i=1}^K \alpha_l  \vect{g}_l  h_{l,i}  s_i + \sqrt{\rho_u} \sum_{i=1}^K  \overline{\vect{h}}_i s_i \nonumber\\
    &\quad +\sum_{l=1}^L  \alpha_l  \vect{g}_l  n_l + \vect{v} \nonumber,  
\end{align}
where $\vect{g}_l\in \mathbb{C}^{M \times 1}$ is the channel from repeater $l$ to the BS, $\overline{\vect{h}}_i\in \mathbb{C}^{M \times 1}$ is the direct channel from UE $i$ to the BS, and $\vect{v}\sim \CN (\vect{0},\sigma_{\rm BS}^2\vect{I}_M) \in \mathbb{C}^{M \times 1}$ is the AWGN. We let $\vect{w}_k \in \mathbb{C}^{M \times 1}$ denote the receive combining vector used to decode the data from UE $k$. The soft estimate of the data symbol $s_k$ becomes 
\begin{align}
  y_k = & 
\underset{\text{Desired signal}}{\underbrace{ \sqrt{\rho_u} \vect{w}^{\Htran}_k  \vect{z}_k  s_k }}  + \underset{\text{Colored noise}}{\underbrace{ \sqrt{\rho_u} \sum_{\underset{i \neq k}{i=1} }^K \vect{w}^{\Htran}_k \vect{z}_i s_i + \sum_{l=1}^L \alpha_l \vect{w}^{\Htran}_k\vect{g}_l  n_l + \vect{w}^{\Htran}_k\vect{v} }},  
\end{align}
where $\vect{z}_i =  \sum_{l=1}^L \alpha_l   h_{l,i} \vect{g}_l + \overline{\vect{h}}_i  $ is the composite channel between the BS and UE $i$ considering both the direct and repeater-assisted paths. We assume perfect channel estimation and the absence of multi-cell interference to establish an upper bound on system performance. The covariance matrix of the colored noise before receive combining can be expressed as 
\begin{equation}
    \vect{C}_k = \sum_{\underset{i \neq k}{i=1} }^K \rho_u  \vect{z}_i \vect{z}^{\Htran}_i + \sum_{l=1}^L \alpha^2_l \vect{g}_l  \vect{g}^{\Htran}_l \sigma^2_{\rm r} + \sigma^2_{\rm BS} \vect{I}_M.
\end{equation}

We can apply the linear minimum mean squared error (LMMSE) combiner at the receiver, where $\vect{w}_k =  \vect{C}^{-1}_k \vect{z}_k$ \cite{bjornson2024introduction}. The SINR for UE $k$ is then given by 
\begin{equation}
    \mathrm{SINR}_k =  \rho_u \vect{z}^{\Htran}_k \vect{C}^{-1}_k \vect{z}_k.
\end{equation}

\section{Repeater Activation Control for Fairness}
\label{sec:max_min}
We aim to maximize the minimum SINR experienced by the UEs to have a fair comparison with the cfmMIMO system. This problem is stated as
\begin{subequations}
  \begin{align}
 & \underset{\{\alpha_1,\ldots,\alpha_L\}}{\text{maximize}}  \quad \min_{k} \,\, \vect{z}^{\Htran}_k \vect{C}^{-1}_k \vect{z}_k  \\ & \textrm{subject to} \nonumber \\ &   \left(\rho_u \|\vect{h}_{l}\|^2  + \sigma^2_{\rm r}  \right) \alpha^2_{l} \leq P_{\mathrm{max}}, \quad \forall l \\ &  \alpha_{l} \leq \alpha_{\mathrm{max}}, \quad \forall l .
\end{align}  
\end{subequations}
This problem can be expressed in epigraph form as 
\begin{subequations}
  \begin{align}
 & \underset{\{\alpha_1,\ldots,\alpha_L, t\}}{\text{maximize}}  \quad t \\ & \textrm{subject to} \nonumber \\ & \vect{z}^{\Htran}_k \vect{C}^{-1}_k \vect{z}_k  \geq t, \quad \forall k \label{eq:opt1:SINR}\\ & \left(\sqrt{\rho_u \|\vect{h}_{l}\|^2  + \sigma^2_{\rm r} } \right) \alpha_{l} \leq \sqrt{P_{\mathrm{max}}}, \quad \forall l \\ &
 \alpha_l \leq \alpha_{\mathrm{max}}, \quad \forall l
\label{eq:opt1:max_pow} .
\end{align}  
\end{subequations}
This is a non-convex problem due to the inverse and quadratic relationships of $\vect{z}_k$ and $\vect{C}_k$. Note that both $\vect{z}_k$  and $\vect{C}_k$ are functions of $\alpha_{l}$. To convexify these relationships, we will use a lower bound on the SINR expressions obtained by the following lemma. 
\vspace{1mm}
\begin{lemma} Let 
$\vect{z}_k(\hat{\boldsymbol{\alpha}})$ and $\vect{C}_k(\hat{\boldsymbol{\alpha}})$ be matrix-valued functions of $\hat{\boldsymbol{\alpha}} = [\alpha_1, \ldots, \alpha_L]^{\Ttran}$. Consider $\vect{z}^{(0)}_{k} = \vect{z}_{k}(\hat{\boldsymbol{\alpha}}^{(0)})$, $\vect{C}^{(0)}_k  = \vect{C}_k(\hat{\boldsymbol{\alpha}}^{(0)})$ are the given values for  $\hat{\boldsymbol{\alpha}} = \hat{\boldsymbol{\alpha}}^{(0)}$ . A lower bound on $\vect{z}^{\Htran}_k \vect{C}^{-1}_k \vect{z}_k$ can be found by 
\begin{align}
    \vect{z}^{\Htran}_k \vect{C}^{-1}_k \vect{z}_k & \geq   \quad 2 \Re\left( {\vect{b}^{(0)}_k}^{\Htran}  \vect{z}_k \right) - \operatorname{tr}\left( \vect{D}^{(0)}_k \vect{C}_k \right) , \quad \forall k  
\end{align}
where $\vect{D}^{(0)}_k = {\vect{C}^{(0)}_k}^{-1} {\vect{z}^{(0)}_k} {\vect{z}^{(0)}_k}^{\Htran}  {\vect{C}^{(0)}_k}^{-1}$, and $ \vect{b}^{(0)}_k = {\vect{C}^{(0)}_k}^{-1}{\vect{z}^{(0)}_k}$.
\label{lemma:lower_bound}
\end{lemma}
\vspace{0.5mm}
\begin{proof}
   As given in  Appendix A of \cite{lower_bound}, $\vect{z}^{\Htran}_k \vect{C}^{-1}_k \vect{z}_k$ is jointly convex in $\vect{z}_k$ and $\vect{C}_k$, where $\vect{z}_k \in \mathbb{C}^{M\times 1}$, and $\vect{C}_k$ is positive definite $\forall k$. A lower bound on $\vect{z}^{\Htran}_k \vect{C}^{-1}_k \vect{z}_k$ can be obtained by a first-order Taylor expansion at any point $\vect{z}^{(0)}_{k}$, $\vect{C}^{(0)}_{k}$:
   \begin{align}
   \vect{z}^{\Htran}_k \vect{C}^{-1}_k \vect{z}_k  & \geq   \quad 
   \operatorname{tr}\left(  {\vect{z}^{(0)}_k}^{\Htran} {\vect{C}^{(0)}_k}^{-1} \vect{z}^{(0)}_k\right)  \nonumber \\ & \quad + 2 \Re\left(  {\vect{z}^{(0)}_k}^{\Htran} {\vect{C}^{(0)}_k}^{-1} \left(\vect{z}_k - \vect{z}^{(0)}_k\right)  \right) \nonumber \\ & \quad - \operatorname{tr}\left( {\vect{C}^{(0)}_k}^{-1} {\vect{z}^{(0)}_k} {\vect{z}^{(0)}_k}^{\Htran}  {\vect{C}^{(0)}_k}^{-1} \left(\vect{C}_k - \vect{C}^{(0)}_k\right) \right), \nonumber \\ &  = \operatorname{tr}\left(  {\vect{z}^{(0)}_k}^{\Htran} {\vect{C}^{(0)}_k}^{-1} \vect{z}^{(0)}_k\right) - 2  \Re\left(  {\vect{z}^{(0)}_k}^{\Htran} {\vect{C}^{(0)}_k}^{-1}  \vect{z}^{(0)}_k  \right) \nonumber \\ & \quad  + \operatorname{tr}\left( {\vect{C}^{(0)}_k}^{-1} {\vect{z}^{(0)}_k} {\vect{z}^{(0)}_k}^{\Htran} \right)  + 2 \Re\left(  {\vect{z}^{(0)}_k}^{\Htran} {\vect{C}^{(0)}_k}^{-1} \vect{z}_k \right) \nonumber \\  & \quad-  \operatorname{tr}\left( {\vect{C}^{(0)}_k}^{-1} {\vect{z}^{(0)}_k} {\vect{z}^{(0)}_k}^{\Htran}  {\vect{C}^{(0)}_k}^{-1} \vect{C}_k   \right)  \nonumber \\ &
   = 2 \Re\left(  {\vect{z}^{(0)}_k}^{\Htran} {\vect{C}^{(0)}_k}^{-1} \vect{z}_k \right)  \nonumber \\ & \quad-  \operatorname{tr}\left( {\vect{C}^{(0)}_k}^{-1} {\vect{z}^{(0)}_k} {\vect{z}^{(0)}_k}^{\Htran}  {\vect{C}^{(0)}_k}^{-1} \vect{C}_k \right).
   \end{align}

  \label{lemma2}
\end{proof}

The lower bound given in Lemma \ref{lemma:lower_bound} can be used to remove the inverse matrices from the optimization problem and separate the intertwined variables. However, this relaxation can only guarantee a Karush-Kuhn-Tucker (KKT) point rather than providing a global optimum \cite{lipp2016variations}. To obtain a convex form in \eqref{eq:opt1:SINR}, we first reformulate $\vect{z}_i$ as $\vect{z}_i = \tilde{\vect{H}}_i \boldsymbol{\alpha}$, where $\tilde{\vect{H}}_i  = \left[ h_{1,i} \vect{g}_1, h_{2,i} \vect{g}_2, \ldots, h_{L,i} \vect{g}_L, \overline{\vect{h}}_i \right] \in \mathbb{C}^{M \times (L+1)}$, and $\boldsymbol{\alpha} = [\hat{\boldsymbol{\alpha}}^{\Ttran}, 1]^{\Ttran}$. In this case, $\vect{C}_k$ can be reformulated as 
\begin{align}
\vect{C}_k = \sum_{\underset{i \neq k}{i=1}}^{K} \rho_u\tilde{\vect{H}}_i \boldsymbol{\alpha} \boldsymbol{\alpha}^{\Ttran} \tilde{\vect{H}}^{\Htran}_i + \sum_{l=1}^L  \alpha^2_l \sigma^2_{\rm r} \vect{G}_l  + \sigma^2_{\rm BS} \vect{I}_M, \label{eq:C} 
\end{align}
where $\vect{G}_l = \vect{g}_l \vect{g}^{\Htran}_l,  \forall l$. By using \eqref{eq:C} and  Lemma 1, we can replace at the $c$th iteration \eqref{eq:opt1:SINR} with 
\begin{align}
2 \Re\left( \left(\vect{b}^{(c)}_k \right)^{\Htran}  \tilde{\vect{H}}_k \boldsymbol{\alpha} \right) -  \sum_{\underset{i \neq k}{i=1}}^{K} \rho_u  \operatorname{tr}\left(   \boldsymbol{\alpha}^{\Ttran} \vect{F}^{(c)}_{i,k} \boldsymbol{\alpha} \right)  - \left(\tilde{\vect{g}}^{(c)}_k \right)^{\Ttran} \boldsymbol{\alpha}_s \nonumber \\ 
- \sigma^2_{\rm BS} \operatorname{tr}\left( \vect{D}^{(c)}_k\right) \geq t, \quad \forall k,
 \label{eq:convexified}
\end{align}

 where $\vect{F}^{(c)}_{i,k} = \tilde{\vect{H}}^{\Htran}_i  \vect{D}^{(c)}_k \tilde{\vect{H}}_i$. $\tilde{\vect{g}}_k^{(c)} = [\tilde
{g}_{1,k}^{(c)}, \ldots, \tilde
{g}_{L,k}^{(c)}]^{\Ttran} $, and    $\tilde
{g}_{l,k}^{(c)} = \sigma^2_{\rm r} \operatorname{tr}(\vect{D}^{(c)}_k \vect{G}_l)$, and $\boldsymbol{\alpha}_s = [\alpha^2_{1}, \ldots, \alpha^2_{L} ]^{\Ttran}$. By replacing \eqref{eq:opt1:SINR} with \eqref{eq:convexified}, the problem becomes convex and can be solved with any convex programming tool. The proposed iterative amplification assignment algorithm is given in Algorithm \ref{alg:1}.

\vspace{2mm}
\begin{algorithm}
	\caption{[\textbf{MaxMin}] Max-min amplification algorithm to guarantee fair SINRs at UEs. } \label{alg:1}
	\begin{algorithmic}[1]
            \State {\bf Input:} $\tilde{\vect{H}}_k$, $\vect{g}_l$, $h_{l,k}$
		\State {\bf Initialization:} Initialize ${\boldsymbol{\alpha}}^{(0)}$  while keeping $\alpha_l^{(0)} \leq \alpha_{\mathrm{max}}, l = 1, \ldots, L$, $\alpha_{L+1}^{(0)} = 1$.  Set the iteration counter to $c=0$. Set the solution accuracy to $\epsilon>0$.
		\While{$ \frac{\| \boldsymbol{\alpha}^{(c)} - \boldsymbol{\alpha}^{(c-1)} \|^2}{\| \boldsymbol{\alpha}^{(c-1)} \|^2}  >\epsilon$}
        \State Calculate $\vect{F}^{(c)}_{i,k}$, $\tilde{\vect{g}}_k^{(c)}$, $\vect{D}^{(c)}_k$, and $ \vect{b}^{(c)}_k$  by using ${\boldsymbol{\alpha}}^{(c)}$.
\State Solve the following problem: 
\begin{subequations}
  \begin{align}
 & \underset{\boldsymbol{\alpha}, t}{\text{maximize}}  \quad t  \\ & \textrm{subject to} \quad \eqref{eq:convexified} \nonumber \\ & 
\left(\sqrt{\rho_u \|\vect{h}_{l}\|^2  + \sigma^2_{\rm r}} \right)\alpha_{l}  \leq \sqrt{P_{\mathrm{max}}} , \quad \forall l \\
& \alpha_{l}  \leq \alpha_{\mathrm{max}} , \quad \forall l.
\end{align} 
\label{eq:final_maxmin}
\end{subequations}
\State Set $\boldsymbol{\alpha}^{(c+1)}$ to the solution of  \eqref{eq:final_maxmin}.
      \State $c \gets c+1$
		\EndWhile
		\State {\bf Output:}  $\boldsymbol{\alpha}$.  
	    \end{algorithmic}
\end{algorithm}
\vspace{2mm}
While repeaters increase the cellular mMIMO performance, they introduce additional power consumption that can increase significantly as the number of repeaters increases. In the following, we address this problem. 

\section{Energy-Efficient Repeater Configuration}
In numerical analysis, we observe that using maximum amplification at repeaters performs well, but this results in wasting unnecessary energy. In this section, we investigate an alternative design that aims to minimize the energy consumption in the repeaters while guaranteeing a certain performance at UEs.  The power consumption of a repeater can be modeled as \cite{repeater_power_consumption}
\begin{equation}
    P_l = 
    \begin{cases} 
    P_{\mathrm{stat}} +\Delta_p  P_{\mathrm{out},l}, & \text{ if } s_l = \text{active}, \\ 
    P_{\mathrm{sleep}}, & \text{ if } s_l = \text{sleep},
    \end{cases}
\end{equation}
where $s_l$ is the state of the repeater, and it can be dynamically adapted in the order of a few hundred milliseconds. $P_{\mathrm{stat}}$ represents the baseline power consumption from, e.g., power supply, oscillators, cooling. $P_{\mathrm{sleep}}$ is the power consumption on the sleeping state. $\Delta_p $ is the slope of the load-dependent power consumption. 

For the sake of simplicity, we will utilize an optimization-based calculation of the output powers of the repeaters (function of $\alpha_l$) and then utilize heuristics to make long-term sleep decisions. To provide an upper bound on the performance of the considered system, we will first assume that the BS controls and adapts the amplification factors of the repeaters in each coherence block while adapting $s_l$ over a long-term horizon.\footnote{The operation in a coherence block can be designed so that first uplink pilot signals are transmitted, followed by downlink transmission, including the repeater control signaling. Finally, UEs can send uplink signals while repeaters are optimally configured.} In the numerical analysis, we will show that the instantaneous control provides insignificant gains, and the solution of the proposed algorithm can be used as long as the large-scale characteristics do not change. 

As the objective function, the total repeater output power, given by $\sum_{l=1}^LP_{{\rm out},l} =\sum_{l=1}^Lc_l^2\alpha_l^2$, can be minimized, where $c_l =  \sqrt{\left(\rho_u \|\vect{h}_{l}\|^2  + \sigma^2_{\rm r}  \right)}, \, \forall l$. However, to promote sparsity and favor repeater deactivation, we replace the $l_2$-norm squared power expression with the $l_1$-norm, resulting in $\sum_{l=1}^Lc_l\alpha_l$. Our numerical experiments indicate that this objective replacement leads to greater power savings.
The instantaneous optimization problem can be described as below.
\begin{subequations}
  \begin{align}
 & \underset{{\alpha_1,\ldots,\alpha_L}}{\text{minimize}}  \quad \sum_{l=1}^L c_l \alpha_l \\ & \textrm{subject to} \quad
\nonumber \\ & 
\vect{z}^{\Htran}_k \vect{C}^{-1}_k \vect{z}_k  \geq \frac{\mathrm{SINR}_{\mathrm{th}, k}}{\rho_u}, \quad \forall k, \label{eq:opt2:SINR}\\ & c_l \alpha_{l} \leq \sqrt{P_{\mathrm{max}}}, \quad \forall l, \\ &
 \alpha_l \leq \alpha_{\mathrm{max}}, \quad \forall l,
\end{align} 
\label{eq:minpow}
\end{subequations}
where $\mathrm{SINR}_{\mathrm{th}, k}$ is the SINR threshold of UE $k$. Lemma \ref{lemma:lower_bound} and the previous convex relaxation analysis for the SINR constraints can be utilized for this problem as well. Then, \eqref{eq:opt2:SINR} becomes identical to \eqref{eq:convexified} when $t$ is replaced by $\frac{\mathrm{SINR}_{\mathrm{th}, k}}{\rho_u}$. However, replacing the optimization variable $t$ with a constant $\frac{\mathrm{SINR}_{\mathrm{th}, k}}{\rho_u}$ might result in infeasibility for a randomly chosen starting point for the CCP algorithm. Therefore, we will take a feasible point pursuit (FPP) approach with the CCP algorithm, where we introduce a nonnegative feasibility parameter $f_k$ for each constraint $k= 1, \ldots, K$, and we will add a regularization term to the objective to reduce these variables to zero iteratively \cite{FPP}. This ensures a feasible start for the algorithm, and the infeasibility of the result can be checked by the convergence of the feasibility parameters. The power minimization problem can be stated as follows: 
\begin{subequations}
  \begin{align}
 & \underset{\alpha_1,\ldots,\alpha_L, f_1, \ldots, f_K}{\text{minimize}}  \quad \sum_{l=1}^L c_l \alpha_l + \lambda \sum_{k=1}^K f_k \\ & \textrm{subject to} \nonumber \\ & 
2 \Re\left( \vect{b}^{(c)^{\Htran}}_k  \tilde{\vect{H}}_k \boldsymbol{\alpha} \right) -  \sum_{\underset{i \neq k}{i=1}}^{K} \rho_u  \operatorname{tr}\left(   \boldsymbol{\alpha}^{\Ttran} \vect{F}^{(c)}_{i,k} \boldsymbol{\alpha} \right) - \tilde{\vect{g}}_k^{(c)^{\Ttran}} \boldsymbol{\alpha}_s   \nonumber \\  &  \vspace{5mm} - \sigma^2_{\rm BS} \operatorname{tr}\left( \vect{D}^{(c)}_k\right) + f_k  \geq \frac{\mathrm{SINR}_{\mathrm{th}, k}}{\rho_u}, \quad \forall k,
\\
&f_k\geq 0, \quad \forall k, \\
&   \alpha_{l}  \leq \min\left\{\alpha_{\mathrm{max}}, \frac{\sqrt{P_{\mathrm{max}} }}{c_l}\right\} , \quad \forall l ,
\end{align} 
\label{eq:final_minpow}
\end{subequations}
where $\lambda$ is the regularization coefficient. 
The proposed algorithm for joint amplification control and sleep decision is given in Algorithm \ref{alg:2}. The algorithm starts by collecting several samples of CSI corresponding to $T$ coherence blocks. Then, it calculates amplification gains for each coherence block. The algorithm then thresholds the amplification coefficients and obtains activation coefficients, $\mathtt{I}_{t,l}$. With these coefficients, we utilize different binary decision functions, which we will explain further in the following section. 
\begin{algorithm}
	\caption{[\textbf{MinPow}] Repeater amplification and sleep decision algorithm to minimize total power consumption. } \label{alg:2}
	\begin{algorithmic}[1]
            \State {\bf Input:} $T$ instances of $\tilde{\vect{H}}_k$, $\vect{g}_l$, $h_{l,k}$.
            \For{$t = 1: T$}
		\State {\bf Initialization:} Initialize ${\boldsymbol{\alpha}}^{(0)}$  while keeping $\alpha_l^{(0)} \leq \alpha_{\mathrm{max}}, l = \{1, \ldots, L\}$, $\alpha_{L+1}^{(0)} = 1$.  Set the iteration counter to $c=0$. Set the solution accuracy to $\epsilon>0$.
		\While{$ \frac{\| \boldsymbol{\alpha}^{(c)} - \boldsymbol{\alpha}^{(c-1)} \|^2}{\| \boldsymbol{\alpha}^{(c-1)} \|^2}  >\epsilon$}
        \State Calculate $\vect{F}^{(c)}_{i,k}$, $\tilde{\vect{g}}_k^{(c)}$, $\vect{D}^{(c)}_k$, and $ \vect{b}^{(c)}_k$  by using ${\boldsymbol{\alpha}}^{(c)}$.
\State Solve \eqref{eq:final_minpow}.
\State Set $\boldsymbol{\alpha}^{(c+1)}$ to the solution of  \eqref{eq:final_maxmin}.
      \State $c \gets c+1$
		\EndWhile
       \State  Set ${\alpha}_{t,l} = {\alpha}^{(c)}_{l}, \quad \forall l$.
       \State Set $\mathtt{I}_{t,l} = \mathbb{I}\left( {\alpha}_{t,l} > \alpha_{\mathrm{thr}} \right), \quad \forall l$.
        \EndFor
        \State $s_l = f_b\left(\{\mathtt{I}_{t,l}\}_{t=1}^T\right)$.
		\State {\bf Output:}  $\alpha_l$, $s_l$, $\forall l$.  
	    \end{algorithmic}
\end{algorithm}

\section{Numerical Analysis}
In this section, we will compare the performance of mMIMO, RA-MIMO, and cfmMIMO. In both mMIMO and cfMIMO, the number of total antennas, either co-located or distributed, is equal to $M=64$.  For RA-MIMO, we consider a simulation area of  $2\text{\,km} \times 2\text{\,km}$ with grid-type repeater deployment, where a mMIMO BS is located in the center. For cfmMIMO, single-antenna access points are distributed across the region in a grid pattern, similar to the repeaters but without a central BS. Unless otherwise stated, the UEs are uniformly distributed in the considered area. For each simulation case, $100$ random UE locations are considered, with $50$ sample channel realizations per UE location. The amplification factors in both Algorithm \ref{alg:1} and Algorithm \ref{alg:2} are initialized with their maximum values since this provides faster convergence and improved performance.  The rest of the simulation parameters are given in Table \ref{tab:simulation_params}. 

\begin{table}[tb!]
    \centering
    \caption{Simulation parameters.}
    \begin{tabular}{|l | c|} \hline
      $M$  &  $64$ \\
     $\rho_u$ & $20\,\text{\,dBm}$ \\
     Repeater height    & $15\,\text{\,m}$ \\
      BS height & $25\,\text{\,m}$ \\
      UE height & $1.5\,\text{\,m}$ \\ 
      Bandwidth  & $20\,\text{\,MHz}$  \\
      Carrier frequency &  $3.5\,\text{\,GHz}$  \\ 
      Channel model & 3GPP UMa \cite{UMA} \\
      K-factor & $9\,\text{\,dB}$ \\
    $P_{\mathrm{max}}$ & $38\,\text{\,dBm}$ \cite{3gpp38106} \\
      Noise figure (repeater \& BS) & $5\,\text{\,dB}$ \\
      Solution accuracy $(\epsilon)$ & $10^{-5}$ \\
       $P_\mathrm{stat}$ & $24.26$\,W \\
       $\Delta_p$ & $2$ \\
       $P_\mathrm{sleep}$ & $4.72$\,W
      \\ \hline
    \end{tabular}
    \label{tab:simulation_params}
\end{table}

\begin{figure}[tb]
    \centering
    \includegraphics[width=\linewidth]{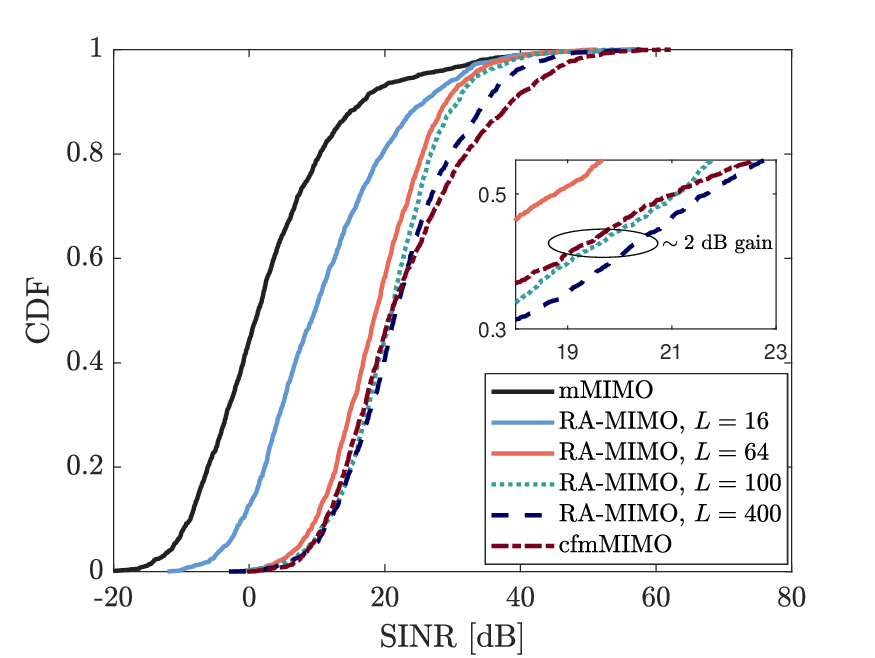}
    \caption{Comparison of RA-MIMO with mMIMO and cfmMIMO. The maximum amplification is chosen for the RA-MIMO results.}
    \label{fig:fig1}
\end{figure}

\begin{table}[tb]
    \centering
        \caption{$\alpha_{\mathrm{max}}$ for different  grid  deployments \cite{larsson2024stability}. }
    \begin{tabular}{|c|l|l|l|l|}  \hline
     $L$  & $16$ & $64$ & $100$ & $400$\\  \hline
       $\alpha_{\mathrm{max}}$ [dB] & $70$ & $58$ & $54$ & $42$  \\  \hline
    \end{tabular}
    \label{tab:alpha}
\end{table}
A simple benchmark can be obtained by assigning the amplification factors of the repeaters to the maximum possible value as $\alpha_{l}  = \min\left\{\alpha_{\mathrm{max}}, \frac{\sqrt{P_{\mathrm{max}} }}{c_l}\right\} , \,\forall l$. As the number of repeaters increases, the allowed $\alpha_{\mathrm{max}}$ decreases as given in Table \ref{tab:alpha}. $K=8$ UEs are uniformly distributed in the considered area. Fig. \ref{fig:fig1} provides the cumulative distribution function (CDF) of the SINR performance of UEs considering RA-MIMO with different numbers of repeaters. The figure shows that repeaters significantly improve the performance of the cell-edge UEs compared to mMIMO. This trend continues as the repeaters are deployed more densely, although $\alpha_{\mathrm{max}}$ is smaller for these cases. An important factor in this trend is the well-interference cancellation performance of mMIMO BS with a high number of antennas. Repeaters act as channel scatters, and both amplify the uplink signals and also provide diversity that benefits interference cancellation at the BS.  When  $L=64$, RA-MIMO is only $1$-$2$\,dB behind of cfmMIMO for most of the UEs, especially for the cell-edge UEs. When $400$ repeaters are deployed, for the cell-edge UEs, RA-MIMO outperforms cfmMIMO with $1$-$2$\,dB, but for the UEs with better channel conditions, cfmMIMO outperforms RA-MIMO. Overall, this figure demonstrates significant potential to improve the cell-edge UE performance of mMIMO BS by deploying repeaters with the maximum amplification mechanism.

\begin{figure}[tb]
    \centering
    \includegraphics[width=\linewidth]{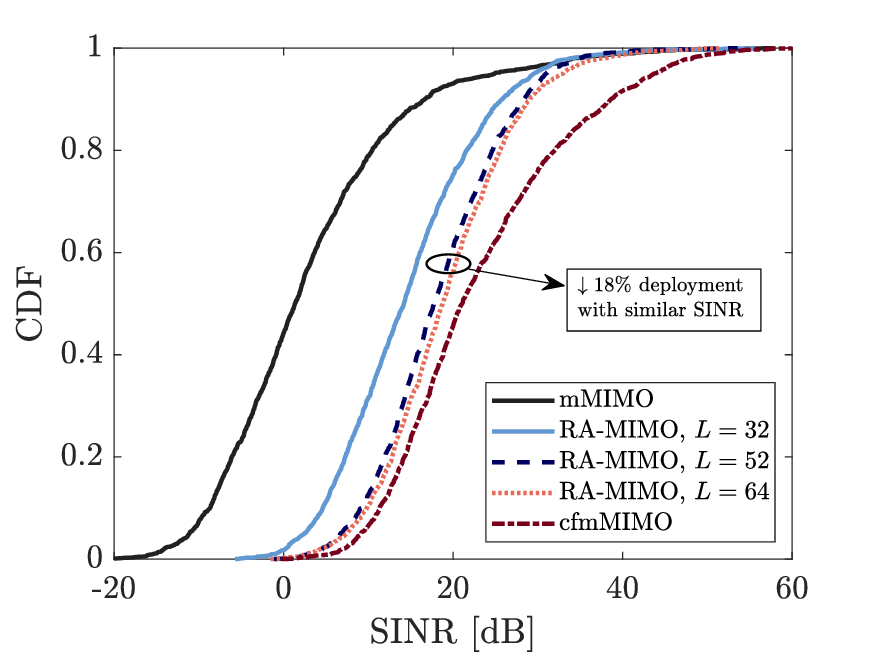}
    \caption{The impact of reducing the repeater deployment on UE SINRs.}
    \label{fig:deployment}
\end{figure}

Reducing the number of repeaters can also be done by removing the repeaters that are close to the BS since mMIMO already performs well for the UEs that are in the cell center. In this way, both the deployment cost and long-term energy cost of the repeaters can be avoided. Fig. \ref{fig:deployment} demonstrates the performance impact of such a scheme, where the repeaters that have a distance shorter than some threshold are removed.\footnote{We consider the same $\alpha_{\mathrm{max}}$ in this figure for different numbers of repeater deployments since the repeaters on the cell edge still have the same inter-repeater distance.} We consider maximum amplification gain at the repeaters for simplicity. As can be seen from the figure, with negligible performance reduction, the deployment of repeaters can be reduced by $18\%$. Reducing the deployment by $50\%$ lowers the SINR performance approximately by $5$\,dB, but still provides significant gains compared to mMIMO.

\begin{figure}[tb]
    \centering
    \includegraphics[width=\linewidth]{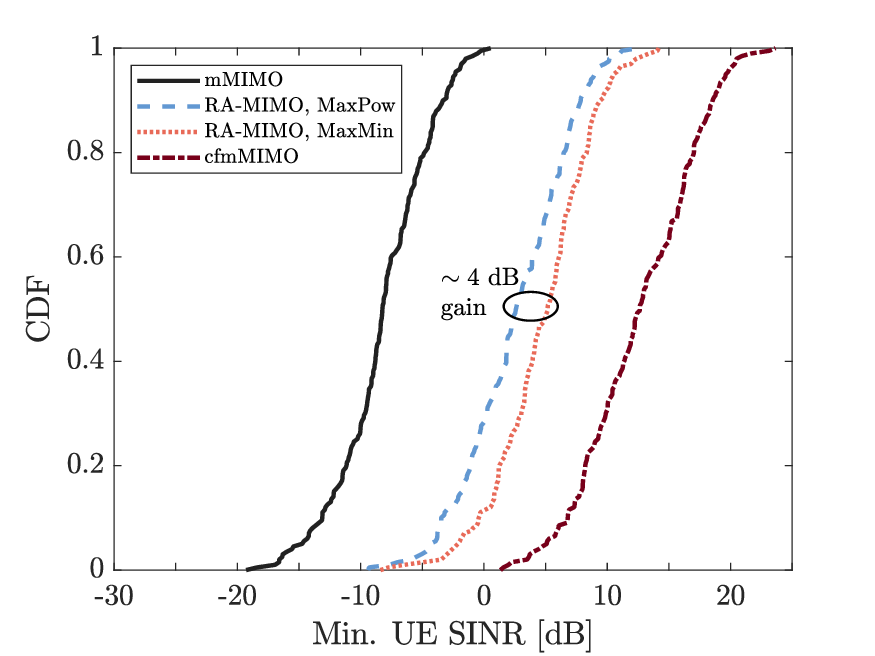}
    \caption{Minimum UE SINR comparison for mMIMO, cfmMIMO, RA-MIMO with maximum amplification (MaxPow) and RA-MIMO with fair control algorithm proposed in Algorithm \ref{alg:1}. UEs are distributed on the cell edge. }
    \label{fig:max_min}
\end{figure}

The maximum amplification might not be beneficial for the setups where the repeaters have low received signal levels. In these cases, they can work as noise amplifiers, where they might predominantly transmit the amplified noise. In such cases, an amplification control becomes necessary, as proposed in Section \ref{sec:max_min}.  To demonstrate this effect, we consider a specific scenario where $K=8$ UEs are distributed on the cell edge, and Cartesian coordinates $x,y \in [1.8, 2]\,\text{km}$. We consider $L=64$ grid deployment for the repeaters. Fig. \ref{fig:max_min} provides the comparison of the minimum SINR among UEs when the repeater amplification is decided with the proposed MaxMin fairness control in Algorithm \ref{alg:1} and with the maximum amplification control. For this specific case, RA-MIMO significantly improves compared to the mMIMO BS case, providing approximately $10\,\text{dB}$ SINR gain to the UEs. Since the UEs are far away from the BS, the RA-MIMO setup is still impacted by the high path loss due to the repeater-BS distance, resulting in $10\,\text{dB}$ performance loss compared to the cfmMIMO setup. The proposed MaxMin control algorithm provides approximately $4\,\text{dB}$ performance gain compared to the maximum power control, demonstrating the sensitivity of the cell-edge UEs to the amplified repeater noise. Since both the received signal through direct UE channels and repeater-assisted channels are weak at the cell edge, the amplified repeater noise reduces the SINR at the BS. However, during our numerical analysis, we observe approximately $1\,\text{dB}$ gain with MaxMin control for the case when UEs are uniformly distributed in the whole region, showing that the sophisticated amplification control mainly helps the cell-edge UEs.

\begin{figure}[tb]
    \centering
\includegraphics[width=\linewidth]{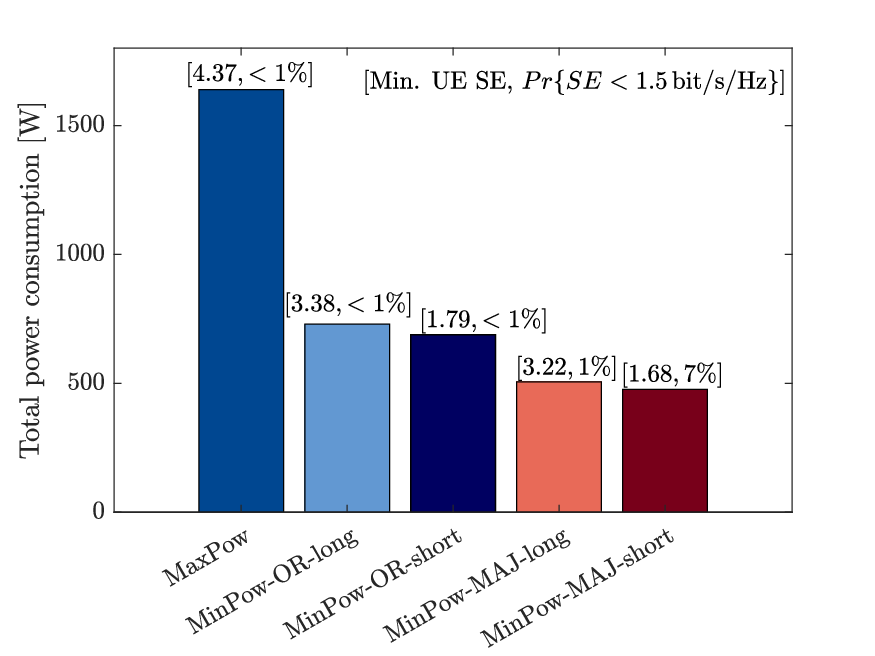}
    \caption{Average of the total power consumption of all repeaters for $K=4$. On $x$-axis, the names of the algorithms are labeled. The numbers in brackets denote the minimum UE SE averaged over different setups and the outage probability from the target SE of $1.5\,\text{bit/s/Hz}$. }
    \label{fig:fig3}
\end{figure}

 Several different heuristics are implemented to minimize power consumption by joint sleep mode and instantaneous amplification control at the repeaters. We first consider two different types of binary functions in Algorithm \ref{alg:2}: (1) OR rule, (2) majority rule. After giving the sleep decisions as in  Algorithm \ref{alg:2}, we consider two different kinds of amplification control: 
\begin{itemize}
    \item Long-term control (MinPow-long): If the repeaters are not in sleep mode, maximum possible amplification is used. In this way, only sleep decisions are sent to the repeaters, requiring significantly less control signaling. 
    \item Short-term control (MinPow-short): If the repeaters are active in each coherence block, the amplification of the repeaters is optimized as in Algorithm \ref{alg:2}. This creates significant synchronization and control signaling with the repeaters but bears the potential of more refined power savings. 
\end{itemize}

Fig. \ref{fig:fig3} shows the power consumption of the considered four different amplification control algorithms for the RA-MIMO system when $K=4$. We set $T=5$, and we consider that UE locations change at every $50$ coherence block.  We run a total of $100$ different UE setups with $50$ coherence blocks. The SE requirement for all UEs is set to $1.5\,\text{bit/s/Hz}$. In each setup, the average minimum SE among UEs is considered the performance metric.  MaxPow is considered the simplest baseline where all repeaters are active and transmitting with full power, as in Fig. \ref{fig:fig1}. The rest of the four are mixed versions of the proposed amplification control algorithms. Long-term decision mechanisms can reduce power consumption by more than $55\%$ while creating less than $1\%$ SE outage for the UEs. Short-term control provides $5\%$ power consumption reduction compared to the long-term control. This reduction can be viewed as insignificant due to the communication overhead and tight synchronization requirement of the repeaters in short-term control. The majority rule reduces the power consumption by $30\%$ compared to the OR rule; however, it increases the outage probability for the short-term decision. Among all algorithms, the majority rule with long-term decision provides the best power/outage trade-off with $70\%$ reduction of power consumption and $1\%$ SE outage.

\begin{table}[tb]
    \centering
        \caption{Run-time comparison of the optimization parts of the proposed algorithms for $K=4$.}
    \begin{tabular}{|l|l|l|l|l|}  \hline
    Algorithm  & Average run-time (sec.) \\  \hline
     MaxPow  & 0.001 \\  
      MaxMin \eqref{eq:final_maxmin}  & 2.14 \\  
      MinPow \eqref{eq:final_minpow} & 5.2  \\ \hline
    \end{tabular}
    \label{tab:run_time}
\end{table}

In the numerical analysis, we implemented the proposed algorithms on MATLAB using CVX and MOSEK. The simulations were executed on a system equipped with an Intel i7 processor.
Table \ref{tab:run_time} compares the run-time of the optimization parts of the proposed algorithms. The total run-time of the MinPow algorithm depends on how many coherence blocks are observed, $T$, to make a sleep decision. The provided time in Table \ref{tab:run_time} considers the solution time based on a single coherence block observation. MaxPow is the simplest algorithm, which employs the maximum allowable amplification within predefined limits; thus, the time required is only for selecting the maximum amplification factor. Since amplification control for MaxMin must be performed in every coherence block, the run-time of this approach is somewhat high for real-time applications. Likewise, when $T = 10$, the run-time of MinPow-long is approximately $53$ seconds, which should ideally be reduced to a few seconds for real-time implementation. However, these algorithms can be used to train neural networks to learn the appropriate policies for different UE distributions, significantly reducing the implementation time. 

\section{Conclusion}
In this paper, we analyzed the uplink performance of a repeater-assisted massive MIMO (RA-MIMO) system and proposed several repeater activation control algorithms, where one targets to enhance cell-edge user performance and another to minimize repeater power consumption. Our numerical results show that RA-MIMO significantly improves the cell-edge performance of conventional mMIMO and closely approaches the performance of cfmMIMO when repeaters are deployed at equal or higher density than cfmMIMO access points, even without amplification control. Additionally, we demonstrate that repeater deployment can be reduced by $18\%$ by eliminating cell-center repeaters with negligible performance loss. Furthermore, our proposed power minimization algorithm, which applies long-term sleep decisions based on the majority rule, reduces total power consumption by $70\%$, highlighting RA-MIMO’s potential as an energy-efficient and scalable solution for next-generation wireless networks. In future work, long-term channel statistics can be utilized to make sleep decisions. Investigating the impact of channel estimation errors and multi-cell interference is also an essential future research direction to showcase the realistic performance of RA-MIMO.

\bibliographystyle{IEEEtran}
\bibliography{IEEEabrv,refs}

\begin{thebibliography}{10}
\providecommand{\url}[1]{#1}
\csname url@samestyle\endcsname
\providecommand{\newblock}{\relax}
\providecommand{\bibinfo}[2]{#2}
\providecommand{\BIBentrySTDinterwordspacing}{\spaceskip=0pt\relax}
\providecommand{\BIBentryALTinterwordstretchfactor}{4}
\providecommand{\BIBentryALTinterwordspacing}{\spaceskip=\fontdimen2\font plus
\BIBentryALTinterwordstretchfactor\fontdimen3\font minus \fontdimen4\font\relax}
\providecommand{\BIBforeignlanguage}[2]{{%
\expandafter\ifx\csname l@#1\endcsname\relax
\typeout{** WARNING: IEEEtran.bst: No hyphenation pattern has been}%
\typeout{** loaded for the language `#1'. Using the pattern for}%
\typeout{** the default language instead.}%
\else
\language=\csname l@#1\endcsname
\fi
#2}}
\providecommand{\BIBdecl}{\relax}
\BIBdecl

\bibitem{massivemimobook}
E.~Bj\"{o}rnson, J.~Hoydis, and L.~Sanguinetti, ``Massive {MIMO} networks: {Spectral}, energy, and hardware efficiency,'' \emph{Foundations and Trends{\textregistered} in Signal Processing}, vol.~11, no. 3-4, pp. 154--655, 2017.

\bibitem{topal2024energy}
O.~A. Topal, {\"O}.~T. Demir, E.~Bj{\"o}rnson, and C.~Cavdar, ``Energy-efficient cell-free massive {MIMO} with wireless fronthaul,'' in \emph{Asilomar Conference on Signals, Systems and Computers}, 2024, arXiv preprint arXiv:2412.02771.

\bibitem{RIS_magazine}
E.~Björnson, {\"O}.~Özdogan, and E.~G. Larsson, ``Reconfigurable intelligent surfaces: {T}hree myths and two critical questions,'' \emph{IEEE Communications Magazine}, vol.~58, no.~12, pp. 90--96, 2020.

\bibitem{aastrom2024ris}
M.~{\AA}str{\"o}m, P.~Gentner, O.~Haliloglu, B.~Makki, and O.~Tageman, ``{RIS} in cellular networks--{C}hallenges and issues,'' \emph{arXiv preprint arXiv:2404.04753}, 2024.

\bibitem{wen2024shaping}
C.-K. Wen, L.-S. Tsai, A.~Shojaeifard, P.-K. Liao, K.-K. Wong, and C.-B. Chae, ``Shaping a smarter electromagnetic landscape: {IAB, NCR, and RIS in 5G} standard and future {6G},'' \emph{IEEE Communications Standards Magazine}, vol.~8, no.~1, pp. 72--78, 2024.

\bibitem{LOS_repeater}
L.-S. Tsai and D.-s. Shiu, ``Capacity scaling and coverage for repeater-aided {MIMO} systems in line-of-sight environments,'' \emph{IEEE Transactions on Wireless Communications}, vol.~9, no.~5, pp. 1617--1627, 2010.

\bibitem{repeater_power_consumption}
A.~Schumacher, R.~Merz, and A.~Burg, ``Increasing cellular network energy efficiency for railway corridors,'' in \emph{Design, Automation \& Test in Europe Conference \& Exhibition (DATE)}, 2022, pp. 1103--1106.

\bibitem{erik_larsson}
S.~Willhammar, H.~Iimori, J.~Vieira, L.~Sundstr{\"o}m, F.~Tufvesson, and E.~G. Larsson, ``Achieving distributed {MIMO} performance with repeater-assisted cellular massive {MIMO},'' \emph{arXiv preprint arXiv:2406.00142}, 2024.

\bibitem{larsson2024stability}
E.~G. Larsson and J.~Bai, ``Stability analysis of interacting wireless repeaters,'' \emph{arXiv preprint arXiv:2405.01074}, 2024.

\bibitem{bjornson2024introduction}
E.~Bj{\"o}rnson and {\"O}.~T. Demir, \emph{Introduction to multiple antenna communications and reconfigurable surfaces}.\hskip 1em plus 0.5em minus 0.4em\relax Now Publishers, Inc., 2024.

\bibitem{lower_bound}
Y.~Liu, Q.~Shi, Q.~Wu, J.~Zhao, and M.~Li, ``Joint node activation, beamforming and phase-shifting control in {IoT} sensor network assisted by reconfigurable intelligent surface,'' \emph{IEEE Transactions on Wireless Communications}, vol.~21, no.~11, pp. 9325--9340, 2022.

\bibitem{lipp2016variations}
T.~Lipp and S.~Boyd, ``Variations and extension of the convex--concave procedure,'' \emph{Optimization and Engineering}, vol.~17, pp. 263--287, 2016.

\bibitem{FPP}
O.~Mehanna, K.~Huang, B.~Gopalakrishnan, A.~Konar, and N.~D. Sidiropoulos, ``Feasible point pursuit and successive approximation of non-convex {QCQPs},'' \emph{IEEE Signal Processing Letters}, vol.~22, no.~7, pp. 804--808, 2015.

\bibitem{UMA}
\BIBentryALTinterwordspacing
{3rd Generation Partnership Project (3GPP)}, ``{Technical Specification Group Radio Access Network; Further Advancements for {E-UTRA} Physical Layer Aspects {(Release 9)}},'' {3GPP}, Tech. Rep. TR 36.814 V9.2.0, Mar. 2017. [Online]. Available: \url{https://portal.3gpp.org/desktopmodules/Specifications/SpecificationDetails.aspx?specificationId=2492}
\BIBentrySTDinterwordspacing

\bibitem{3gpp38106}
\BIBentryALTinterwordspacing
------, ``{5G; NR Repeater Radio Transmission and Reception (Release 18)},'' {3GPP}, Tech. Rep. TS 38.106 V18.4.0, May. 2024. [Online]. Available: \url{https://portal.3gpp.org/desktopmodules/Specifications/SpecificationDetails.aspx?specificationId=3620}
\BIBentrySTDinterwordspacing

\end{thebibliography}

\end{document}